\documentclass[preprint,nofootinbib]{revtex4}

\usepackage{graphicx}

\usepackage{amsmath}

\usepackage{xcolor}

\begin{document}

\title{Relativistic entanglement in muon decay}

\author{S. Carneiro$^{1}$\footnote{Corresponding author\\saulocarneiro@on.br\\ORCID: 0000-0001-7098-383X} and F.C. Sobrinho$^{2}$}

\affiliation{$^1$Observat\'orio Nacional, 20921-400, Rio de Janeiro, RJ, Brazil\\$^2$Instituto de F\'isica, Universidade de S\~ao Paulo, 05508-090, S\~ao Paulo, SP, Brazil}

\date{\today}

\begin{abstract}

We discuss the time evolution of quantum entanglement in the presence of non-collapsing interactions. In particular, the entanglement between the products of a muon decay in a magnetic field is revisited. It results from angular momentum conservation and leads to an anomaly in the measured muon $g$ factor in precise agreement with that reported by the Brookhaven and Fermilab experiments, alleviating the present tension between data-driven and lattice theoretical values.

\vline


\end{abstract}

\maketitle

\section{Introduction}

Among all the strange facets of the quantum world, entanglement is certainly the most profound evidence of the singular behaviour of quantum systems, especially of the inherent non-locality of their phenomenology \cite{entanglement}. It is usually manifest in destructive measurements, when the collapse of the system wave function leads to correlations between observations that, from a classical viewpoint, are not causally connected -- for example, of the polarisation of particles resulting from the decay of a scalar system. Although the orthodox interpretation of this phenomenon invokes the very process of measurement, the primary basis for any realistic description of non-locality is
the character of the wave function as an attribute of the whole system.

In this way, entanglement would, in principle, also be manifest in unitarity preserving processes, whenever the evolutions of different parts of a system are not separable. Let us consider, for example, a two-particle system with wave function $\Psi(s_1,s_2)$, where $s$ is any observable attributed to each particle, e.g. their spins and momenta. If we want to know the probability distribution associated to the first particle, a summation must be done on the possible values of the observables of the second,
\begin{equation}
    |\psi(s_1)|^2 = \sum_{s_2} |\Psi(s_1,s_2)|^2.
\end{equation}
Let us imagine that $s_2$ changes with time due to some interaction with an external field. The wave functions $\Psi(s_1,s_2)$, in this case, do not constitute a stationary basis, 
and the expansion above leads to a probability distribution for $s_1$ that evolves with time.

A context where such a correlation occurs is that of a weak decay in a magnetic field, where the probability distribution of an emitting particle has a non-separable dependence on the spins of the others. The precession of the latter around the field leads to a time modulation in the former when we trace over the spins. In relativistic decays only the initial and final states can be defined and we cannot really follow such a modulation. Nevertheless, its signature would be statistically visible in an ensemble of decays with random flying times to the emitted particles detection. A concrete example of this type will be discussed here.

\section{The muon decay}

Consider a relativistic, horizontally polarised muon decaying into a positron and two neutrinos inside a vertical, uniform magnetic field {\bf $B$}. In the muon frame the preferred direction of positron emission is aligned with the muon spin \cite{Tiomno} and this fact was ingeniously used for a precise measurement of the muon Larmor frequency \cite{PRL,prd,2006}. The time modulation in the positron distribution is a direct measurement of the muon magnetic dipole precession. Now, let us evaluate the effect of the neutrinos precession on the result, reminding that the positron probability amplitude involves a summation over the neutrinos helicities.

Tracing over the neutrinos degrees of freedom, we lose information about the angular momentum carried by them, which leads to an uncertainty in the angular momentum of the parent muon \cite{americano}. Let us consider a neutrino emitted in the $x$-direction, orthogonal to the magnetic field that points along the $z$-direction, and that its polarisation is initially longitudinal. For a massless neutrino the polarisation would remain longitudinal and its angular momentum in the lab frame would be given by its spin $s_x$. 

A massive neutrino, however, presents both left and right-handed helicities. Tracing over them leaves the positron in a state with no well defined angular momentum\footnote{We are not taking into accounting the positron helicities, which are not measured or traced over and play no role in the muon Larmor frequency determination.}.
The neutrino precesses around the magnetic field, acquiring an angular momentum in the $y$-direction. Its value in the lab frame can be derived with the help of a Lorentz transformation of the angular momentum tensor, and it is given by
\begin{equation} \label{L}
    L_{y} = \gamma_{\nu} \left(s_y - v_{\nu} L^{'03} \right),
\end{equation}
where $v_{\nu}$ is the neutrino velocity, $\gamma_{\nu}$ is its Lorentz factor, and
\begin{equation}
    L^{'03} = t'p'_z - E' z'
\end{equation}
is the $03$ component of the angular momentum tensor in the neutrino rest frame, being ${\bf p}'$, $E'$ and ${\bf r}'$ its linear momentum, energy and position at a time $t'$, respectively \cite{Landau}.
As the neutrino is freely moving, $L^{'03}$ is constant. Therefore, the angular momentum acquired after an infinitesimal time interval $dt$ will be
\begin{equation}
    \delta L_y = \gamma_{\nu} \frac{d s_y}{dt} dt = -\gamma_{\nu} \mu_{\nu} B dt,
\end{equation}
where $\mu_{\nu}$ is the neutrino magnetic dipole and we have used $d{\bf s}/dt = {\bf \mu_{\nu}} \times {\bf B}$.

The corresponding uncertainty in the muon angular momentum due to neutrino tracing is twice this value, because, in the lab frame, the neutrino and anti-neutrino resulting from the decay are emitted in the same forward direction. Applying to the muon the same reasoning as above, we have
\begin{equation}
    2\delta L_y = -\gamma_{\mu} \delta \mu_{\mu} B dt,
\end{equation}
where $\delta \mu_{\mu}$ is the uncertainty in the muon magnetic moment, and $\gamma_{\mu}$ is its Lorentz factor. Equating these two results we obtain
\begin{equation} \label{Eq.1}
    \delta \mu_{\mu} \gamma_{\mu} = 2 \mu_{\nu} \gamma_{\nu}.
\end{equation}
In other words, we can state that the correction in the muon proper frequency is twice the neutrino proper frequency. This correlation between proper periods -- the only meaningful in a Lorentz invariant theory -- was {\it postulated} in Ref.~\cite{BJP} in order to justify Eq.~(\ref{Eq.1}). Here we have shown that such a postulate is not needed, being a consequence of angular momentum conservation.

In the simplest case where massive Dirac neutrinos are described by the inclusion of right-handed singlets in the standard model, their magnetic moment was precisely calculated, giving the leading order term \cite{mu,PDG}
\begin{equation}
    \mu_{\nu} = \frac{3eG_Fm_{\nu}}{8\sqrt{2}\pi^2},
\end{equation}
where $e$ is the elementary charge, $G_F$ is the Fermi constant and $m_{\nu}$ is the neutrino mass. Eq.~(\ref{Eq.1}) can then be rewritten as
\begin{equation} \label{2}
    \delta a_{\mu} \equiv \left( \frac{2m_{\mu}}{e} \right) \delta \mu_{\mu} = \left( \frac{3m_{\mu}^2G_F}{4\sqrt{2}\pi^2} \right) f \approx 7.2 f \times 10^{-9},
\end{equation}
where $\delta a_{\mu}$ is the correction in the muon magnetic anomaly, $m_{\mu}$ is the muon mass, and
\begin{equation}
f \equiv \frac{2m_{\nu}\gamma_{\nu}}{m_{\mu}\gamma_{\mu}}
\end{equation}
is the fraction of the muon energy carried by the neutrinos in the decay. Its distribution is peaked around $f \approx 0.35$ \cite{2006}, leading to $\delta a_{\mu} \approx 2.5 \times 10^{-9}$, in remarkable accordance with the Brookhaven and Fermilab experiments.

\begin{figure}
    \centering
    \includegraphics[scale=.59]{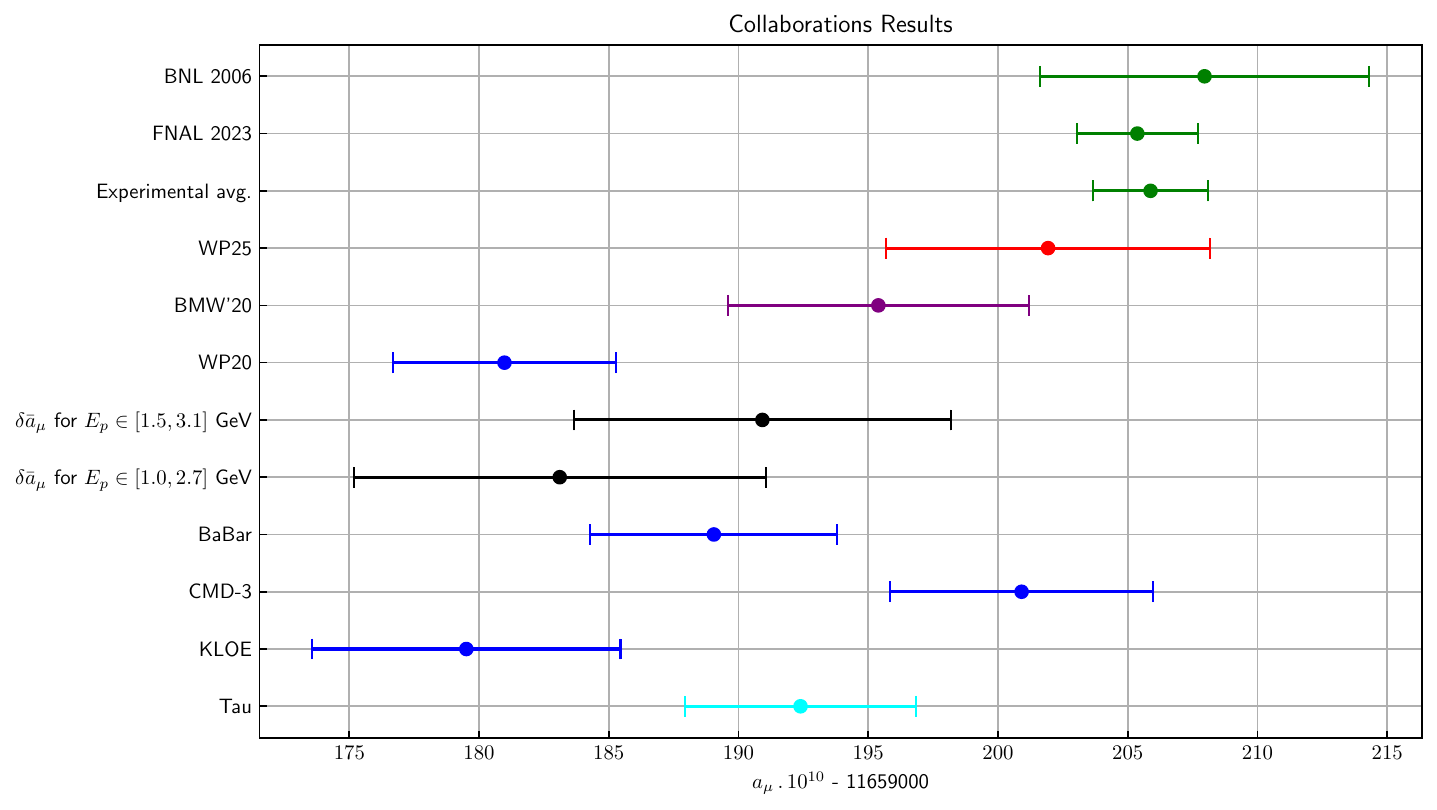}
    \caption{Comparison between our corrected magnetic anomaly $a_{\mu}^{\text{exp}} - \delta \bar{a}_{\mu}$ (black bars) and those derived from data-driven \cite{white} (blue and cyan bars) and lattice \cite{lattice,lattice2} (purple and red bars) calculations. Our two confidence intervals correspond to averaging $\delta a_{\mu}$ in two different intervals of positron energy. The green bars correspond to (uncorrected) measured values.}
   \end{figure}
 
The discrepancy found between the data-driven theoretical value reported in \cite{white} and observations is $\delta a_{\mu} \approx (2.14 \pm 0.50)$ ppm \cite{PRL,prd}, subsequently verified with a higher precision \cite{newMuon,verynew}. From Eq.~(\ref{2}), the anomaly average on the positron energy in the interval $1$ GeV $< E_p < 2.7$ GeV,\footnote{For $E_p < 1$ GeV the asymmetry becomes negative and the error bars are too large for $E_p < 1.5$ GeV, while the highest experimental sensitivity is reached for $E_p \approx 2.6$ GeV.} weighted with the positron asymmetry, is $\delta a_{\mu} \approx (1.96 \pm 0.68)$ ppm \cite{BJP}. If we, instead, average the anomaly in the observed range $1.5$ GeV $< E_p < 3.1$ GeV, we obtain $\delta a_{\mu} \approx (1.28 \pm 0.62)$ ppm. Our corrected values for $a_{\mu}$ are shown in Fig.~1, where they are compared to the values derived from data-driven \cite{white} and lattice calculations \cite{lattice,lattice2}.\footnote{{\color{black}For details on the asymmetry weighted average, see \cite{BJP}.}}

The most recent report of the muon $g-2$ theory initiative \cite{lattice2} has found a better concordance between the theoretical and observed $g$-factors. Nevertheless, it is based on lattice results, and it does not explain the tension with the data-driven results, which still persists.  We can see that, when using the most precise observed range of positron energy, our corrected magnetic anomaly presents $1\sigma$ agreement with both data-driven and lattice calculations, attenuating in this way the current tension between the two sets.

Equation (\ref{Eq.1}) predicts a dependence of the anomaly on the neutrinos energy that can be confirmed or ruled out once precise binned data are released. A preliminary analysis, with the use of the BNL data and weighting $\delta a_{\mu}$ with the positron asymmetry, has shown a little better best-fit as compared to an energy-independent anomaly \cite{BJP}.

{\color{black} \section{Discussion}

A common criticism of the mechanism proposed here is that any subsequent evolution of the unobserved neutrino sector must disappear after tracing over the neutrino degrees of freedom, so that no imprint can survive in the measured positron signal. In our view, that objection is not decisive, because it assumes from the outset that the standard premises of the trace-invariance argument are valid in the present relativistic setting.

That assumption is nontrivial. On one hand, the decay considered here is a relativistic three-body process with correlated final-state momenta, and the laboratory description involves boosted states in an external magnetic field. In such systems, the separation between helicity and momentum degrees of freedom is subtle, since Lorentz boosts induce momentum-dependent Wigner rotations. It is well known that reduced spin density matrices obtained by tracing over momentum are not Lorentz invariant, and that the corresponding subsystem structure is not as straightforward as in finite-dimensional nonrelativistic systems \cite{1,2}.

Furthermore, in the Muon g-2 experiment the observable is constructed through time spectra, energy thresholds and energy-dependent asymmetry weighting \cite{3}. Operationally, this is a coarse-grained, detector-defined observable rather than an ideal measurement on the full final-state Hilbert space. Therefore, the criticism would have to show not merely that partial trace is basis independent, but that, after relativistic kinematics, momentum correlations and detector acceptance are taken into account, the relevant positron and neutrino sectors admit the clean tensor-product decomposition required by the usual non-signaling theorem.

On the other hand, in the present relativistic context, the relevant operation is not an abstract partial trace performed at the decay vertex over a fixed neutrino Hilbert space. In relativistic quantum theory, only asymptotic {\it in} and {\it out} states have an unambiguous physical meaning, and no observable time evolution is defined between them in general \cite{Landau 4}. The relevant out state here is therefore the detector-defined asymptotic final state selected by the measurement process. The measured signal is reconstructed from sub-ensembles of events selected by detection time, energy and acceptance. In this setting, the final neutrino sector contains two helicity channels and, after tracing over them, the positron is left in a mixed angular state whose statistical weights are time dependent. Since the muon anomaly is inferred from the time modulation of detected positrons, this induces an additional modulation with characteristic frequency determined by the neutrino Larmor precession or, if unresolved, an effective shift or broadening of the fitted anomaly.

Our claim is not that relativistic subsystem subtleties prove the effect proposed here, nor that any retroactive or noncausal influence is transmitted from the neutrino sector back to the muon or positron sector. The point is narrower: if part of the angular momentum structure of the decay remains encoded in the unobserved neutrino sector, then tracing over that sector may correspond not to a signal transmitted between subsystems, but to an incompleteness in the mapping between the measured positron modulation and the inferred muon anomaly (as quantified in Eq. (6)). In this sense, the proposal does not conflict with the non-signaling theorem; rather, it questions whether the standard inference from the positron modulation to the muon anomaly is fully closed once the angular degrees of freedom carried by the unobserved neutrinos are taken into account.}

\section{Concluding comments}

The observation of quantum entanglement in high energy experiments involving the decay of particles is already a reality \cite{graziano,Atlas}. They will certainly challenge the current developments of a relativistic theory for this intriguing phenomenon, which is still far from being completely understood \cite{Peres,relativity1,relativity2}. Special Relativity brings to the problem of entanglement new and interesting complications that have not found fully satisfactory solutions. It is not easy to conciliate the non-local character of the wave function with the local causality respected by classical observers. A typical example is given in Alice and Bob type experiments, where a correlation is found between two subsequent measurements -- for instance of the polarisations of two particles that once formed an unpolarised system -- performed by observers with no causal connection. The temporal order between these two events depends on the reference frame, since we have a space-like interval \cite{Aharanov}. This is an important challenge for the interpretation of the wave function collapse as resulting from the interaction with the apparatus of measurement.

In this paper we have discussed an aspect that may appear -- and perhaps have already appeared -- in the high energy context, namely the manifestation of entanglement in unitarity conserving processes like the Larmor precession. In the concrete example of the muon decay, entanglement appears from tracing over the neutrinos time-dependent helicities, leaving an uncertainty in the angular momentum of the parent muon. The resulting correction in the muon magnetic moment presents a precise agreement with the anomaly reported in the Brookhaven and Fermilab experiments. A dependence of the observed anomaly on the positron energy is also predicted, which was corroborated by the Brookhaven dataset and may be confirmed or falsified when more precise energy-binned data are available.

\section*{Acknowledgements}

The authors are thankful to D. Boyanovsky, P. Girotti, P. C. de Holanda, F. Navarra, A. Saa and D. Vanzella for useful discussions. FCS thanks FAPESP (SP, Brazil) for his PhD grant 2024/19103-9. SC is partially supported by CNPq (Brazil) with grant 308518/2023-3.




















{\color{black} \bibitem{1} A. Peres, P. F. Scudo, and D. R. Terno, Phys. Rev. Lett. {\bf 88}, 230402 (2002).

\bibitem{2} H. Lee and I. Kim, J. Math. Phys. {\bf 63}, 012201 (2022).

\bibitem{3} D. P. Aguillard {\it et al.} [Muon g-2 Collaboration], Phys. Rev. {\bf D110}, 032009 (2024).

\bibitem{Landau 4} V. B. Berestetiskii, E. M. Lifshitz and L. P. Pitaeviskii, {\it Quantum Electrodynamics} (Pergamon, 1982), sec.~1.}

\bibitem{graziano} D. Babusci {\it et al.} [KLOE-2 Collaboration], JHEP {\bf 04}, 059 (2022).

\bibitem{Atlas} G. Aad {\it et al.} [ATLAS Collaboration], Nature {\bf 633}, 542 (2024).

\bibitem{Peres} A. Peres and D. R. Terno, Rev. Mod. Phys. {\bf 76}, 93 (2004).

\bibitem{relativity1} N. Friis, R. A. Bertlmann, M. Huber and B. C. Hiesmayr, Phys. Rev. {\bf A81}, 042114 (2010).

\bibitem{relativity2} J. Fan and X. Li, Phys. Rev. {\bf D97}, 016011 (2018).

\bibitem{Aharanov} Y. Aharonov and J. Tollaksen, arXiv: 0706.1232 [quant-ph].

\end{document}